\documentclass[conference]{IEEEtran}
\usepackage[utf8]{inputenc}
\usepackage[linesnumbered,ruled,vlined]{algorithm2e}
\usepackage{amssymb}
\usepackage{todonotes}
\usepackage{multirow}
\usepackage{comment}
\usepackage{times}
\usepackage[english]{babel}
\usepackage{booktabs}
\usepackage{url}
\usepackage{hyperref}

\newcommand{\Orca}{\textsc{Orca}}

\begin{document}

\title{ORCA -- a Benchmark for Data Web Crawlers}


\author{\IEEEauthorblockN{Michael R{\"o}der}
\IEEEauthorblockA{Department of Computer Science\\Paderborn University, Germany;\\
Institute for Applied Informatics\\
Leipzig, Germany\\
michael.roeder@upb.de}
\and
\IEEEauthorblockN{Geraldo de Souza Jr.\\ and Denis Kuchelev \\ and Abdelmoneim Amer Desouki}
\IEEEauthorblockA{Department of Computer Science\\Paderborn University, Germany}
\and
\IEEEauthorblockN{Axel-Cyrille Ngonga Ngomo}
\IEEEauthorblockA{Department of Computer Science\\Paderborn University, Germany;\\
Institute for Applied Informatics\\
Leipzig, Germany\\
axel.ngonga@upb.de}}

\maketitle

\begin{abstract}
The number of RDF knowledge graphs available on the Web grows constantly. Gathering these graphs at large scale for downstream applications hence requires the use of crawlers. Although Data Web crawlers exist, and general Web crawlers could be adapted to focus on the Data Web, there is currently no benchmark to fairly evaluate their performance. 
Our work closes this gap by presenting the \Orca{} benchmark. \Orca{} generates a synthetic Data Web, which is decoupled from the original Web and enables a fair and repeatable comparison of Data Web crawlers. Our evaluations show that \Orca{} can be used to reveal the different advantages and disadvantages of existing crawlers. The benchmark is open-source and available at \url{https://w3id.org/dice-research/orca}.
\end{abstract}


\section{Introduction}

The number of RDF knowledge graphs (KGs) available on the Web has grown continuously over recent years.\footnote{For an example, see \url{https://lod-cloud.net/}.} 
These KGs are provided through means ranging from SPARQL endpoints over simple dump files to information embedded in HTML pages. Data Web crawlers are employed to discover and make use of such KGs \cite{ldspider2010}.
With the growing amount of available KGs on the web, these crawlers become more important, e.g., in government-funded projects.\footnote{For example, the projects OPAL (\url{http://projekt-opal.de/}) and LIMBO (\url{https://www.limbo-project.org/}) funded by the German government.}
The efficiency and effectiveness of such crawlers are typically evaluated by crawling the Web for a set amount of time while measuring different performance indicators such as the number of requests performed by the crawler~\cite{Heydon99mercator:a,ldspider2010}. While this kind of experiment can be performed for a crawler at a given point in time, the experiments are virtually impossible to repeat and thus, hard to compare with similar experiments. This is due to several factors, including primarily the fact that the Web is an ever-changing, evolving network of single, partly unreliable nodes. 
Another influence is the geographical location of the machine on which the crawler is executed. For example, geo-blocking can have an influence on the shape of the crawled network. Executing the same crawler on the same hardware might also lead to different evaluation results when various internet service providers offering different connections and bandwidths are used.  
In addition, the ground truth is not known in such experiments. Since the content of the complete Web is unknown, it is hard to measure the effectiveness of a crawler, i.e., its ability to retrieve relevant data.

To overcome these limitations, we propose \Orca{}---a benchmark for Web Data Crawlers. The basic idea of \Orca{} is to alleviate the limitations of current benchmarking approaches by (1) generating a synthetic Data Web and (2) comparing the performance of crawlers within this controlled environment. The generation of the synthetic Web is based on statistics gathered from a sample of the real Data Web. The deterministic generation process implemented by our approach ensures that crawlers are benchmarked in a repeatable and comparable way. 

This paper has four main contributions. 
\begin{enumerate}
    \item We provide an approach to generate a synthetic Data Web. 
    \item Based on this generator, we present our second contribution, \Orca{}, the first extensible FAIR benchmark for Data Web crawlers, which can measure the efficiency and effectiveness of crawlers in a comparable and repeatable way. 
    \item We present the first direct comparison of 2 Data Web crawlers in a repeatable setup. 
    \item We show how \Orca{} can be used to evaluate the politeness of a crawler, i.e., whether it abides by the Robots Exclusion Protocol~\cite{robots2019}.
\end{enumerate}

The rest of the paper is organised as follows. Section~\ref{sec:relatedWork} presents related work while Section~\ref{sec:prerequisites} defines prerequisites. In Section~\ref{sec:approach}, we describe the approach and its implementation. The experiments and their results are presented in Section~\ref{sec:evaluation} and discussed in Section~\ref{sec:discussion}. 
We conclude the paper with Section~\ref{sec:conclusion}.

\section{Related Work}
\label{sec:relatedWork}

We separate our overview of the related work into two parts. First, we present related publications regarding crawlers and their evaluations. Note that due to the limited space, we mainly focus on Data Web crawlers. Second, we present a brief overview of related work, with statistics regarding the Semantic Web.

\paragraph{Crawlers and their Evaluation.}


The Mercator Web Crawler~\cite{Heydon99mercator:a} is an example of a general Web crawler. The authors describe the major components of a scalable Web crawler and discuss design alternatives. The evaluation of the crawler comprises an 8-day run, which has been compared to similar runs of the Google and Internet Archive crawlers. As performance metrics, the number of HTTP requests performed in a certain time period, and the download rate (in both documents per second and bytes per second) are used. 
\cite{srinivasan2005topicaleval} presents an evaluation framework for comparing topical crawlers, i.e., crawlers that are searching for web pages of a certain topic. It relies on a given topic hierarchy and the real web, which makes it susceptible for the aforementioned drawbacks.
The BUbiNG crawler~\cite{boldi2018bubing} has been evaluated relying on the real web as well as a simulation. This simulation was carried out by using a proxy that generated synthetic HTML pages. However, the authors do not give further details about the structure of the simulated web.

A crawler focusing on structured data is presented in~\cite{Harth2006}. It comprises a 5-step pipeline and converts structured data formats like XHTML or RSS into RDF. The evaluation is based on experiments in which the authors crawl 100k randomly selected URIs. To the best of our knowledge, the crawler is not available as open source project. 
In~\cite{Hogan2011Thesis,Hogan2011:SWSE}, a distributed crawler is described, which is used to index resources for the Semantic Web Search Engine. 
In the evaluation, different configurations of the crawler 
are compared, based on the time the crawler needs to crawl a given amount of seed URIs. To the best of our knowledge, the crawler is not available as open-source project. In~\cite{lodlaundromat2014}, the authors present the LOD Laundromat---an approach to download, parse, clean, analyse and republish RDF datasets. The tool relies on a given list of seed URLs and comes with a robust parsing algorithm for various RDF serialisations. In~\cite{lodalot}, the authors use the LOD Laundromat to provide a dump file comprising 650K datasets and more than 28 billion triples.
Two open-source crawlers for the Data Web are LDSpider~\cite{ldspider2010} and Squirrel~\cite{roeder2020squirrel}. Both are described in detail in Section~\ref{sec:benchmarkedCrawlers}.

Apache Nutch~\cite{khare2004nutch} is an open-source Web crawler.
However, the only available plugin for processing RDF stems from 2007, relies on an out-dated crawler version and was not working during our evaluation.\footnote{\url{https://issues.apache.org/jira/browse/NUTCH-460}}

\paragraph{The Data Web.}
\label{sec:lod-sota}


There are several publications analysing the Web of data that are relevant for our work, since we use their insights to generate a synthetic Data Web. The Linked Open Data (LOD) Cloud diagram project periodically generates diagrams representing the LOD Cloud and has grown from 12 datasets in 2007 to more than 1200 datasets in 2019.\footnote{\url{https://lod-cloud.net/}}
These datasets are entered manually, require a minimum size and must be connected to at least one other dataset in the diagram. Other approaches for analysing the Data Web are based on the automatic gathering of datasets. LODStats~\cite{lodstats2012,lodstats2016} collects statistical data about more than 9\,000 RDF datasets gathered from a dataset catalogue.\footnote{The dataset catalogue is \url{http://thedatahub.org}.}
In a similar way, \cite{schmachtenberg2014} uses the LDSpider crawler~\cite{ldspider2010} to crawl datasets in the Web. 
In~\cite{Hogan2012}, the authors gather and analyse 3.985 million open RDF documents from 778 different domains regarding their conformity to Linked Data best practices.  
The authors of \cite{sparql-discoverability2013} compare different methods to identify SPARQL endpoints on the Web and 
suggest that most SPARQL endpoints can be found using dataset catalogues. \cite{schmachtenberg2014} confirms this finding by pointing out that only 14.69\% of the crawled datasets provide VoID metadata. 
In \cite{bizer2013deployment}, the authors analyse the adoption of the different technologies, i.e., RDFa~\cite{rdfa2015}, Microdata~\cite{microdata2014} and Microformats and crawl 3 billion HTML pages to this end. None of these works target a benchmark for Data Web crawlers. We address this research gap with the work. 

\section{Preliminaries}
\label{sec:prerequisites}

\paragraph{Data Web Crawler.}
\label{sec:crawlerDef}

Throughout the rest of the paper, we model a crawler as a program that is able to (1) download Web resources, (2) extract information from these resources and (3) identify the addresses of other Web resources within the extracted information. It will use these (potentially previously unknown) addresses to start with step 1 again in an autonomous way. A Data Web crawler is a crawler that extracts RDF triples from Web resources. Note that this definition excludes programs like the LOD Laundromat~\cite{lodlaundromat2014}, which download and parse a given list of Web resources without performing the third step.

\paragraph{Crawlable Graph.}
\label{sec:crawlableGraph}

Let $\mathbb G=(\mathbb V, \mathbb E)$ be a directed graph.  $\mathbb V=\{v_1, v_2, \ldots\}$ is the set of nodes of the graph. $\mathbb E \subseteq \mathbb V^2$ is the set of edges of $\mathbb G$. Given an edge $e=(u,v)$, we call $u$ the source node and $v$ the target node of $e$. Let $S \subseteq \mathbb V$ be a set of seed nodes. We call a graph $\mathbb G$ \emph{crawlable w.r.t.} $S$ iff all nodes $v \in \mathbb V$ can be reached from nodes $u \in S$ in a finite number of steps by traversing the edges of the graph following their direction. 
A special case of crawlable graphs are graphs that are crawlable w.r.t. a singleton $S=\{v_{\epsilon}\}$. We call $v_{\epsilon}$ an \emph{entrance node} of such graphs.



\paragraph{Data Web Analysis.}
\label{sec:lod-stats}
The Data Web comprises servers of varying complexity. The types of nodes in this portion of the Web include simple file servers offering their data as dump files, Web servers able to dereference single RDF URIs or to serve HTML web pages with embedded structured data, and SPARQL endpoints that are able to handle complex queries. 
Hence, we define the different types of nodes in the synthetic Data Web that is to be generated and used for the benchmark: (1) \textbf{Dump file node.} This node comprises an HTTP server offering RDF data as a single dump file. In its current implementation, \Orca{} randomly chooses one of the following RDF serialisations: RDF/XML, Notation 3, N-Triples, or Turtle. Additionally, the file might be compressed with one of three available compression algorithms---ZIP, Gzip or bzip2.\footnote{Details regarding the compressions can be found at \url{https://pkware.cachefly.net/Webdocs/APPNOTE/APPNOTE-6.3.5.TXT}, \url{https://www.gnu.org/software/gzip/} and \url{http://sourceware.org/bzip2/}, respectively.} (2) \textbf{Dereferencing node.} This node comprises an HTTP server and answers requests to single RDF resources by sending all triples of its RDF graph that have the requested resource as subject. The server offers all serialisations supported by Apache Jena.\footnote{\url{https://jena.apache.org/}} When a request is received, the serialisation is chosen based on the HTTP \texttt{Accept} header sent by the crawler. The complete list of serialisations supported by \Orca{} can be seen in Table~\ref{tab:serializations}. (3) \textbf{SPARQL endpoint.} This node offers an API, which can be used to query the RDF data using SPARQL via HTTP.\footnote{In it's current version, \Orca{} uses Virtuoso for this type of node.}
(4) \textbf{RDFa.} This node offers HTML web pages via HTTP. The web pages contain structured data as RDFa. In its current version, the RDFa node relies on RDFa 1.0 and RDFa 1.1 test cases for HTML and XHTML of an existing RDFa test suite.\footnote{\url{http://rdfa.info/test-suite/}}
(5) \textbf{CKAN.} CKAN is a dataset catalogue containing meta data about datasets.\footnote{\url{https://ckan.org/}} It offers human-readable HTML pages and an API that can be used to query the catalogue content.


\paragraph{Robots Exclusion Protocol.}
The Robots Exclusion Protocol allows the definition of rules for bots like crawlers~\cite{robots2019}. The draft of the standard defines two rules---\texttt{allow} and \texttt{disallow}. They allow or disallow access to a certain path on a domain, respectively. The rules are defined in a \texttt{robots.txt} file, which is typically hosted directly under the domain in which the rules have been defined. Although additional rules are not covered by the standard, the standard allows the addition of lines. Some domain owners and crawlers make use of a \texttt{Crawl-delay} instruction to define how much delay a crawler should have between its requests to this single Web server.\footnote{Examples are Bing (\url{https://blogs.bing.com/Webmaster/2012/05/03/to-crawl-or-not-to-crawl-that-is-bingbots-question/}) and Yandex (\url{https://yandex.com/support/Webmaster/controlling-robot/robots-txt.html}).}

\section{Approach}
\label{sec:approach}

The main idea behind \Orca{} is to ensure the comparable evaluation of crawlers by creating a local, synthetic Data Web. The benchmarked crawler is initialised with a set of seed nodes of this synthetic cloud and asked to crawl the complete cloud. Since the cloud is generated, the benchmark knows exactly which triples are expected to be crawled and can measure the completeness of the crawl and the speed of the crawler. Since the cloud generation is deterministic, a previously used cloud can be recreated for benchmarking another crawler, ensuring that evaluation results are comparable if the experiments are executed on the same hardware. 

It should be noted that our approach relies on a local, synthetic Web comprising containerized servers This differs to the proxy-based approach of~\cite{boldi2018bubing} in which a single proxy implementation simulates the web. This allows us to use real-world implementations of servers (e.g., SPARQL endpoints) and eases future additions of further node types.

In the following, we describe the benchmark in detail. We begin by explaining the cloud generation in Section~\ref{sec:cloudGeneration}. An overview of the implementation and its details is given in Section~\ref{sec:implementation}.

\subsection{Cloud Generation}
\label{sec:cloudGeneration}

Since the synthetically generated Data Web will be used to benchmark a Data Web crawler, we generate it as a crawlable graph w.r.t. a set of seed nodes $S$ as defined in Section~\ref{sec:crawlableGraph}. 
The generation of the synthetic Web can be separated into three steps---(1) Generating the single nodes of the cloud, (2) generating the node graph, i.e., the edges between the nodes, and (3) generating the RDF data contained in the single nodes.

\paragraph{Node Generation.}

The set of nodes $U$ is generated by virtue of types selected from the list of available types in Section~\ref{sec:lod-stats}. The number of nodes in the synthetic Web ($\nu$) and the distribution of node types are user-defined parameters of the benchmark. The node generation process makes sure that at least one node is created for each type with an amount $>0$ in the configuration. 
Formally, let $\Psi=\{\psi_1, \psi_2, \ldots\}$ be the set of node types and $\Psi_u \subseteq \Psi$ be the set of node types to be generated. To ensure that every type occurs at least once, the generation of the first $|\Psi_u|$ nodes of $U$ is deterministic and ensures every type in $\Psi_u$ is generated.
The remaining types are assigned using a seeded random model based on the user-defined distribution until $|U|=\nu$.

\paragraph{Node Graph Generation.}

\begin{table}[tb]
    \centering
    \caption{Connectivity matrix $\mathcal C$ used for the experiments.}
    \label{tab:connectivityMatrix}
    \begin{tabular}{@{}c|ccccc@{}}
    \toprule
         from $\backslash$ to & Deref. & Dump file & SPARQL & CKAN & RDFa \\
         \midrule
         Deref.      & 1 & 1 & 1 & 1 & 1 \\
         Dump file   & 1 & 1 & 1 & 1 & 1 \\ 
         SPARQL      & 1 & 1 & 1 & 1 & 1 \\
         CKAN        & 0 & 1 & 1 & 1 & 1 \\
         RDFa        & 1 & 1 & 1 & 1 & 1 \\
    \bottomrule
    \end{tabular}
\end{table}

In the real-world Data Web, connections (i.e., edges) between instances of certain node types are unlikely. For example, an open data portal is very likely to point to dump files, SPARQL endpoints or even other open data portals. However, it is very unlikely that it points to a single RDF resource, i.e., to a server which dereferences the URI of the resource. To model this distribution, we introduce a connectivity matrix. Let $\mathcal C$ be a $|\Psi|\times|\Psi|$ matrix. $c_{ij}=1$ means that edges from nodes of type $\psi_i$ to nodes of type $\psi_j$ are allowed. Otherwise, $c_{ij}=0$. An example of such a connectivity matrix is given in Table~\ref{tab:connectivityMatrix} and will be used throughout the paper. It allows all connections except the example mentioned above.

The algorithm that generates the node graph takes the matrix $\mathcal C$, the previously created list of typed nodes, and the user-configured average node degree as input. It starts with the first $|\Psi_u|$ nodes of $U$ and creates connections between them. For these initial nodes, all connections allowed in $\mathcal C$ are created.
This initial graph is extended step-wise by adding the other nodes from $U$. In each step, the next node from the list is added to the graph. The outgoing edges of the new node are added using a weighted sampling over the nodes that are permissible from the new node according to $\mathcal C$. Since the Web is known to be a scale-free network, the weights are the in-degrees of the nodes following the Barab{\'{a}}si-Albert model for scale-free networks~\cite{albert2002}. In the same way, a similar number of connections to the new node are generated. 


After generating the node graph, a set of seed nodes $S$ has to be generated to make the graph crawlable as described in Section~\ref{sec:crawlableGraph}. This search is equivalent to the set cover problem~\cite{karp1972}. Hence, searching for the smallest set of seed nodes would be NP-hard. Thus, we use a greedy solution which takes $\mathbb V$ and $ \mathbb E$ of the generated node graph as input. It starts with defining all nodes as unmarked nodes. After this, the first unmarked node is added to $S$. A breadth-first search starts from this node and marks all reachable nodes. After that, the steps are repeated by choosing another unmarked node until all nodes are marked, i.e., all nodes are reachable from the nodes in $S$.

\paragraph{RDF Data Generation.}

The benchmark can work with any RDF data generator. However, it comes with a simple generator that ensures the crawlability of the graph.
The generator can be configured with three parameters: the average number of triples per graph ($\tau$), the distribution of the sizes of the single graphs and the average degree of the RDF resources ($d$) in the graph. In its current version, \Orca{} offers a simple approach that statically assigns the given average size to every RDF graph. However, this can be changed to use any other distribution. 

Let $\mathcal G=(R, P, L, T)$ be an RDF graph, where $R=\{r_1, r_2, \ldots\}$ is the set of URI resources of the graph, $P=\{p_1, p_2, \ldots\}$ is the set of properties, $L$ is the set of external URI resources, i.e., resources belonging to a different RDF graph, with $R \cap L = \varnothing$ and $T=\{t_1, t_2, \ldots \}$ being the set of triples of the RDF graph where each triple has the form $t_j=\{(s_j,p_j,o_j) | s_j \in R, p_j \in P, o_j \in (R \cup L)\}$.\footnote{Note that this simplified definition of an RDF graph omits the existence of literals and blank nodes.} $T$ can be separated into two subsets $T=T_{i} \cup T_{o}$. The set of graph-internal triples $T_{i}$ comprises triples with objects $o_j \in R$. In contrast, the set of outgoing triples $T_{o}$ (a.k.a. link set) contains only triples with external resources as objects ($o_j \in L$). Further, let $d$ be the average node degree of the resources, i.e., the number of triples a resource is part of.

Like the node graph, each created RDF graph has to be crawlable w.r.t. a set of resources. For the RDF graphs, we implemented an algorithm based on the Barab{\'{a}}si-Albert model for scale-free networks~\cite{albert2002}. The algorithm guarantees that all resources within the generated RDF graph can be reached from the first resource it generates. As defined in Section~\ref{sec:crawlableGraph}, this resource can be used later on as entrance node by all other RDF graphs which must generate links to this graph. 

Let $\tau$ be the RDF graph size that has been determined based on the chosen parameters. Based on the previously created node graph, the number of outgoing edges $\tau_o=|T_{o}|$ as well as their objects, i.e., the set of external URI resources $L$, are known. Algorithm~\ref{alg:rdfGraph} takes $\tau_{i} = \tau - \tau_{o}$ together with the average degree $d$ and a generated set of properties $P$ as input to generate an initial version of graph $\mathcal G$. 
The loop (lines~\ref{line:nodeAddStart}--\ref{line:addNode}) adds new resources to the graph until the number of necessary triples has been reached. For each new resource $r_n$, a URI is generated (line~\ref{line:uriCreation}) before it is connected to the existing resources of the graph. After that, the degree of the new resource $d_r$ is drawn from a uniform distribution in the range $[1,2d]$ (line~\ref{line:drawDegree}). The $d_r$ resources $r_n$ will be connected to are chosen based on their degree, i.e., the higher the degree of a resource, the higher the probability that it will be chosen for a new connection. The result of this step is the set $R_c$ with $|R_c|=d_r$. For each of these resources, a direction of the newly added triple is chosen. Since the graph needs to be crawlable, the algorithm chooses the first triple to be pointing to the newly resourced node. This ensures that all resources can be reached, starting from the first resource of the graph. For every other triple, the decision is based on a Bernoulli distribution with a probability of $\frac{0.5d_r-1}{d_r-1}$ being a triple that has the new node as an object  (line~\ref{line:choosingDirection}). 
Based on the chosen direction, the new triple is created with a property that is randomly drawn from the property set $P$ (lines~\ref{line:incomingTriple} and~\ref{line:outgoingTriple}).

\begin{algorithm}[t]
\DontPrintSemicolon
\SetKwFunction{generateResource}{genResource}\SetKwFunction{degree}{degree}\SetKwFunction{drawDegree}{drawDegree}
\SetKwFunction{bernoulli}{bernoulli}\SetKwFunction{generateTriple}{genTriple}\SetKwFunction{draw}{draw}\SetKwFunction{drawFromDegreeDist}{drawFromDegreeDist}
\SetKwFunction{Union}{Union}\SetKwFunction{FindCompress}{FindCompress}
\SetKwInOut{Input}{Input}\SetKwInOut{Output}{Output}
\Input{$\tau_{i}, P, d$}
\Output{$\mathcal G$}
\BlankLine
$E,T_{i}\leftarrow\{\}$\;
$R\leftarrow\{r_1\}$\;
$d_r\leftarrow$\drawDegree{d}\;\label{line:drawDegree}
\While{$|T_{i}| < \tau_{i}$}{\label{line:nodeAddStart}
$r_n\leftarrow$ \generateResource{$|R|$}\;\label{line:uriCreation}
\While{$(|T_{i}| < \tau_{i}) \&\& ($\degree{$r_n$}$ < d_r)$}{
$R_c \leftarrow $ \drawFromDegreeDist{$R,T_{i}$}\;
\For{$r_c \in R_c$}{
\eIf{$($\degree{$r_n$}$ == 0) || ($\bernoulli{$\frac{0.5d_r-1}{d_r-1}$}$)$}{\label{line:choosingDirection}
  $T_{i} \leftarrow T_{i} \cup \{$\generateTriple{$r_c,$\draw{$P$}$,r_n$}$\}$\;\label{line:incomingTriple}
}{
  $T_{i} \leftarrow T_{i} \cup \{$\generateTriple{$r_n,$\draw{$P$}$,r_c$}$\}$\;\label{line:outgoingTriple}
}
}
}
$R \leftarrow R \cup \{r_n\}$\label{line:addNode}
}
$\mathcal G \leftarrow \{R, P, \varnothing, T_{i}\}$\;
\caption{Initial RDF graph generation}
\label{alg:rdfGraph}
\end{algorithm}

After the initial version of the RDF graph is generated, the outgoing edges of $T_{o}$ are created. For each link to another dataset, a triple is generated by drawing a node from the graph as subject, drawing a property from $P$ as predicate and the given external node as object. Both---$T_{o}$ and $L$---are added to $\mathcal G$ to finish the RDF graph.

\paragraph{URI Generation.}

Every resource of the generated RDF graphs needs to have a URI. To make sure that a crawler can use the URIs during the crawling process, the URIs of the resources are generated depending on the type of node hosting the RDF dataset. All URIs contain the host name. The dump file node URIs contain the file extension representing the format of the file (the RDF serialization and the compression). If a resource of the SPARQL node is used in another generated RDF graph (i.e., to create a link to the SPARQL node), the URL of the SPARQL API is used instead of a resorce URI.  In a similar way, the links to the CKAN nodes are created by pointing to the CKAN's Web interface.

\subsection{Implementation}
\label{sec:implementation}

\paragraph{Overview.}

\Orca{} is a benchmark built upon the HOBBIT benchmarking platform~\cite{hobbit2019}.\footnote{\url{https://github.com/hobbit-project/platform}} This FAIR benchmarking platform allows Big Linked Data systems to be benchmarked in a distributed environment. It relies on the Docker\footnote{\url{https://www.docker.com/}} container technology to encapsulate the single components of the benchmark and the benchmarked system. 

\begin{figure}[tb]
    \centering
    \includegraphics[width=0.48\textwidth]{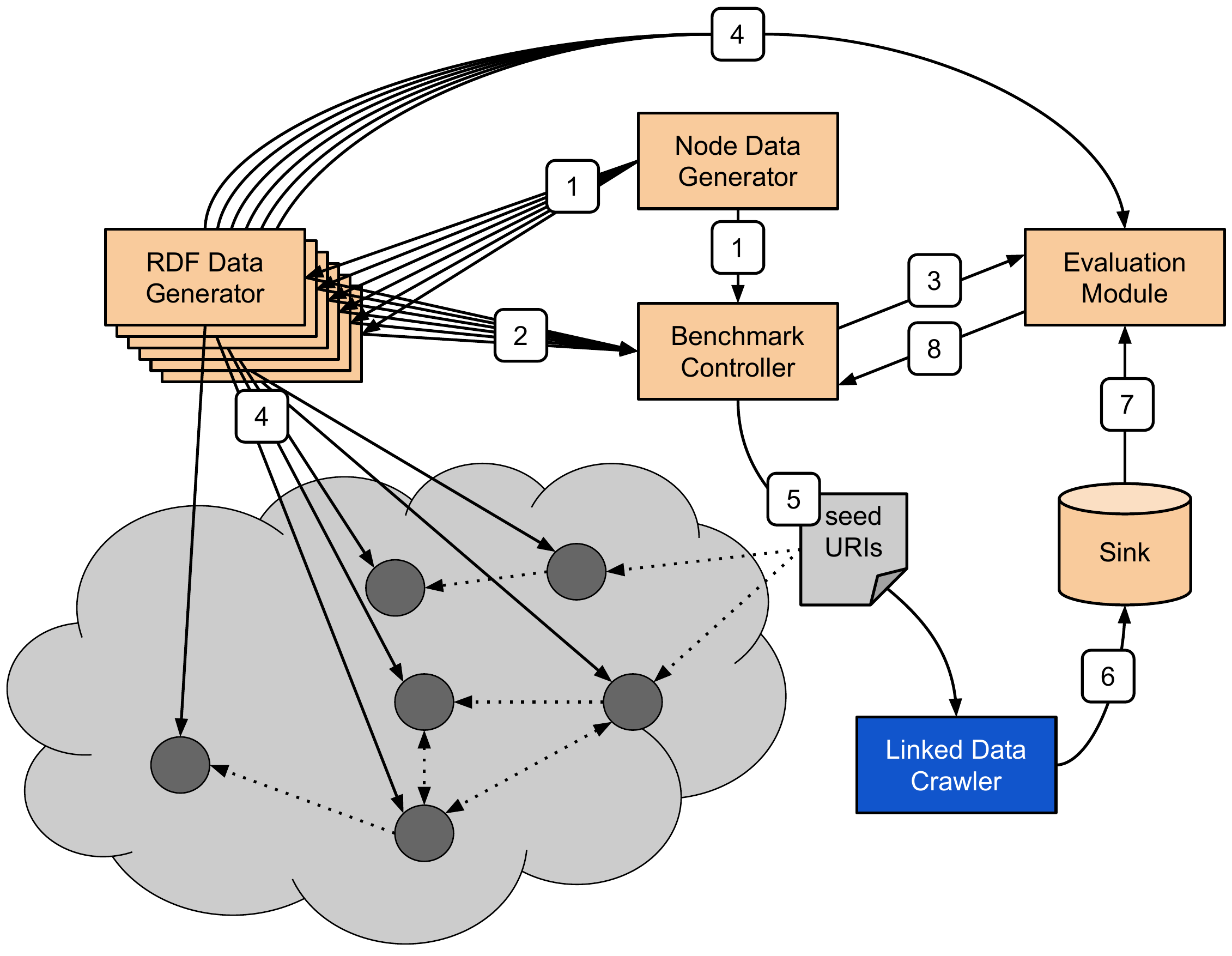}
    \caption{Overview of the Benchmark components and the flow of data. Orange: Benchmark components; Grey: Synthetic Data Web generated by the benchmark; Dark blue: The benchmarked crawler; Solid arrows: Flow of data; Dotted arrows: Links between RDF datasets; Numbers indicate the order of steps.}
    \label{fig:overview}
\end{figure}

We adapted the suggested design of a benchmark described in~\cite{hobbit2019} to implement \Orca{}. The benchmark comprises a benchmark controller, data generators, an evaluation module, a triple store and several nodes that form the synthetic Data Web. The benchmark controller is the central control unit of the benchmark. It is created by the HOBBIT platform, receives the configuration defined by the user and manages the other containers that are part of the benchmark.
Figure~\ref{fig:overview} gives an overview of the benchmark components, the data flow and the single steps of the workflow. The workflow itself can be separated into 4 phases---creation, generation, crawling and evaluation. When the benchmark is started, the benchmark controller creates the other containers of the benchmark.\footnote{The benchmarked crawler is created by the HOBBIT platform as described in~\cite{hobbit2019}.} During this creation phase, the benchmark controller chooses the types of nodes that will be part of the synthetic Data Web, based on the parameters configured by the user. The Docker images of the chosen node types are started together with an RDF data generator container for each node that will create the data for the node. Additionally, a node data generator, a triple store and the evaluation module are started. The node data generator will generate the node graph. The triple store serves as a sink for the benchmarked Linked Data crawler during the crawling phase while the evaluation module will evaluate the crawled data during the evaluation phase.

After the initial creation, the graph generation phase is started. This phase can be separated into two steps---initial generation and linking. During the first step, each RDF data generator creates an RDF graph for its Web node. In most cases, this is done using the algorithm described in Section~\ref{sec:cloudGeneration}. For data portal and RDFa nodes, the generation process differs. The portal nodes solely use the information to which other nodes they have to be linked to, i.e., each linked node is inserted as a dataset into the data portal node's database. The RDFa node relies on an already existing test suite.\footnote{\url{http://rdfa.info/test-suite/}} It generates an HTML page that refers to the single tests and to all other connected nodes using RDFa. The node data generator creates the node graph as described in Section~\ref{sec:cloudGeneration}. After this initial generation step, the node graph is sent to the benchmark controller and all RDF data generators (Step 1 in Figure~\ref{fig:overview}). This provides the RDF data generators with the information to which other nodes their RDF graph should be linked. Subsequently, the RDF data generators send their metadata to each other and the benchmark controller (Step 2). This provides the data generators with the necessary data to create links to the entrance nodes of other RDF datasets during the linking step. Additionally, the benchmark controller forwards the collected metadata to the evaluation module and the nodes in the cloud (Step 3).\footnote{The submission to the cloud nodes has been omitted in the figure.}
At the end of the generation phase, the generated RDF graphs are forwarded to the single nodes and the evaluation module (Step 4). The generation phase ends as soon as all nodes have 
processed the received data.

After the generation phase is finished and the HOBBIT platform signals that the crawler has initialised itself, the benchmark controller submits the seed URIs to the crawler (Step 5). This starts the crawling process in which the crawler sends all downloaded RDF data to its sink (Step 6). When the crawler finishes its crawling---i.e., all given URIs and all URIs found in the crawled RDF data have been crawled---the crawler terminates and the crawling phase ends.

During the evaluation phase, the evaluation module measures the recall of the crawler by checking whether the RDF graphs generated by the data generators can be found in the sink (Step 7). The result of this evaluation is sent to the benchmark controller, which adds further data and results of the benchmarking process (Step 8). This can include data that has been gathered from the single nodes of the cloud, e.g., access times. After this, the final results are forwarded to the HOBBIT platform.

\section{Evaluation}
\label{sec:evaluation}

For evaluating Data Web crawlers, we use three different experiments. The first experiment uses all available node types to generate the synthetic Data Web and mainly focuses on the recall of the benchmarked crawlers. The second experiment uses a simpler Web to measure efficiency. The last experiment checks whether the crawlers abide by the Robots Exclusion Protocol. 


For all experiments, the online instance of HOBBIT is used. It is deployed on a cluster with 3 servers that are solely used by the benchmark and 3 servers that are available for the crawler.\footnote{Each of the servers has 16 cores with Hyperthreading and 256 GB RAM. The details of the hardware setup that underlies the HOBBIT platform can be found at \url{https://hobbit-project.github.io/master\#hardware-of-the-cluster}.}

\subsection{Benchmarked crawlers}
\label{sec:benchmarkedCrawlers}

\begin{table*}[htb]
\centering
    \caption{The RDF serialisations and compressions supported by \Orca{} and the two benchmarked crawlers. (\checkmark) marks serialisations in \Orca{} that are not used within HTML files or for generating dump nodes for the synthetic Linked Data Web. X marks serialisations that are listed as processible by a crawler but were not working during our evaluation.}
    \label{tab:serializations}
    \begin{tabular}{@{}lccccccccccp{0.2cm}cccp{0.2cm}ccc@{}}
    \toprule
        & \multicolumn{10}{c}{\textbf{RDF Serialisations}}
        && \multicolumn{3}{c}{\textbf{Comp.}} && \multicolumn{3}{c}{\textbf{HTML}} \\
    \cmidrule{2-11}\cmidrule{13-15}\cmidrule{17-19}
        & 
        \rotatebox{90}{RDF/XML} &
        \rotatebox{90}{RDF/JSON} &
        \rotatebox{90}{Turtle} &
        \rotatebox{90}{N-Triples} &
        \rotatebox{90}{N-Quads} &
        \rotatebox{90}{Notation 3} &
        \rotatebox{90}{JSON-LD} &
        \rotatebox{90}{TriG} &
        \rotatebox{90}{TriX} &
        \rotatebox{90}{HDT} &&
        \rotatebox{90}{ZIP} &
        \rotatebox{90}{Gzip} &
        \rotatebox{90}{bzip2} &&
        \rotatebox{90}{RDFa} &
        \rotatebox{90}{Microdata} &
        \rotatebox{90}{Microformat} \\
        \midrule
        \Orca{}  & 
        {\checkmark} & 
        {(\checkmark)} & 
        {\checkmark} & 
        {\checkmark} & 
        {(\checkmark)} & 
        {\checkmark} & 
        {(\checkmark)} & 
        {(\checkmark)} & 
        {(\checkmark)} & 
        {--} &&
        {\checkmark} & 
        {\checkmark} & 
        {\checkmark} &&
        {\checkmark} & 
        {--}& 
        {--}\\
        \midrule
        LDSpider & \checkmark & --         & \checkmark & X          & \checkmark & \checkmark & \checkmark &  --        &  --        &   --       && --         & --         & --         && \checkmark & \checkmark & \checkmark \\
        Squirrel & \checkmark & \checkmark & \checkmark & \checkmark & \checkmark & \checkmark & \checkmark & \checkmark & \checkmark & \checkmark && \checkmark & \checkmark & \checkmark && \checkmark & \checkmark & \checkmark \\
    \bottomrule
    \end{tabular}
\end{table*}


To the best of our knowledge, there are only two working open-source Data Web crawlers available---LDSpider and Squirrel. Other crawlers are either not available as open-source project or web crawlers without the ability to process RDF data. Table~\ref{tab:serializations} shows the RDF serialisations supported by LDSpider and Squirrel in comparison to \Orca{}.

LDSpider~\cite{ldspider2010} is an open-source Linked Data crawler that has been used in several publications to crawl data from the Web~\cite{schmachtenberg2014}.\footnote{\url{https://github.com/ldspider/ldspider}} Following the documentation, it is able to process triples serialised as RDF/XML, N3 and Turtle. Additionally, it supports Apache Any23, which can be used to parse RDFa, Microdata, Microformats and JSON-LD.\footnote{\url{https://any23.apache.org/}} The crawler uses multiple threads to process several URIs in parallel. It offers different crawling strategies---a classic breadth-first strategy (BFS), and a load-balancing strategy (LBS). The latter tries to crawl a given number of URIs as fast as possible by sending parallel requests to different domains without overloading the server of a single domain. The crawler can write its results either to files or send them to a triple store. For our experiments, we dockerized LDSpider and implemented a system adapter to make it compatible with the HOBBIT platform. We created several LDSpider instances with different configurations. LDSpider (T1), (T8), (T16) and (T32) use BFS and 1, 8, 16 or 32 threads, respectively. During our first experiments, we encountered issues with LDSpiders' SPARQL client, which was not storing the crawled data in the provided triple store. To achieve a fair comparison of the crawlers, we extended our system adapter to implement our own SPARQL client, used LDSpider's file sink to get the output of the crawling process, and sent file triples to the benchmark sink. These instances of LDSpider are marked with the addition ``FS''. Additionally, we configured the LDSpider instance (T32,FS,LSB), which makes use of the load-balancing strategy to compare the two strategies offered by the crawler.

Squirrel~\cite{roeder2020squirrel} is an open-source, distributed Linked Data crawler, which uses Docker containers to distribute its components. It crawls resources in a similar way to the LSB strategy of LDSpider, by grouping URIs based on their domain and assigning the created URI sets to its workers. Following the documentation, it supports all RDF serialisations implemented by Apache Jena. It uses Apache Any23 to parse Microdata, Microformats and the Semargl parser for RDFa.\footnote{\url{https://github.com/semarglproject/semargl}} Furthermore, it supports the crawling of HDT dump files~\cite{hdt}, SPARQL endpoints and open data portals. The latter includes the crawling of CKAN portals and configuration of a scraper that can extract information from HTML pages following predefined rules. For compressed dump files, Squirrel implements a recursive decompression strategy for ZIP, Gzip, bzip2 and tar files. For our experiments, we implemented an adapter for the Squirrel crawler. Squirrel (W1), (W3), (W9) and (W18) are instances of the crawler using 1, 3, 9 or 18 worker instances, respectively.\footnote{Since the HOBBIT cluster assigns 3 servers for the benchmarked crawler, we use multiples of 3 for the number of workers.}

\subsection{Data Web Crawling}

The first experiment simulates a real-world Data Web and focuses on the effectiveness of the crawlers, i.e., the amount of correct triples they retrieve. To define the distribution of node types, we analyse the download URLs of the LODStats~\cite{lodstats2012,lodstats2016} dump from 2016. Based on this analysis, the generated cloud comprises 100 nodes with 40\% dump file nodes, 30\% SPARQL nodes, 21\% dereferencing nodes and 5\% CKAN nodes. 
4\% are RDFa nodes to represent the 1 billion RDFa triples identified by~\cite{bizer2013deployment} in comparison to the 28 billion RDF triples gathered from the semantic web by~\cite{lodalot}. Following~\cite{Hogan2012}, the average degree of each node is set to 20. For each node, the RDF graph generation creates 1000 triples with an average degree of 9 triples per resource.\footnote{The dataset size is roughly the average document size from~\cite{Hogan2012} (after excluding an outlier). The average degree is derived from the statistics of DBpedia, Freebase, OpenCyc, Wikidata and Yago from~\cite{faerber2018}.} Based on our LODStats analysis, 30\% of the dump file nodes use one of the available compression algorithms for the dump file. The usage of \texttt{robots.txt} files is disabled.
Since LDSpider does not support the crawling of SPARQL endpoints, data catalogues like CKAN, or compressed dump files, we expect LDSpider to achieve a lower Recall than Squirrel. The results of the experiment are listed in Table~\ref{tab:results}.\footnote{The detailed results can be seen at \url{https://w3id.org/hobbit/experiments\#1584544940477,1584544956511,1584544971478,1585403645660,1584545072279,1585230107697,1584962226404,1584962243223,1585574894994,1585574924888,1585532668155,1585574716469}.}



\begin{table*}[tb]
    \centering
    \caption{Results of the Data Web crawling and efficiency experiments.}
    \label{tab:results}
    \begin{tabular}{@{}lp{0cm}cp{0.1cm}rp{0.2cm}cp{0.1cm}rp{0.1cm}rp{0.1cm}r@{}}
    \toprule
        \multicolumn{1}{c}{\textbf{\multirow{3}{*}{Crawler}}} 
        && \multicolumn{3}{c}{\textbf{Data Web}} 
        && \multicolumn{7}{c}{\textbf{Efficiency}}
        \\
        \cmidrule{3-5}\cmidrule{7-13}
        && \multicolumn{1}{c}{\textbf{Micro}} && \multicolumn{1}{c}{\textbf{Runtime}}
        && \multicolumn{1}{c}{\textbf{Micro}} && \multicolumn{1}{c}{\textbf{Runtime}}
        && \multicolumn{1}{c}{\textbf{CPU}} && \multicolumn{1}{c}{\textbf{RAM}}
        \\
        && \multicolumn{1}{c}{\textbf{Recall}} && \multicolumn{1}{c}{(in s)}
        && \multicolumn{1}{c}{\textbf{Recall}} && \multicolumn{1}{c}{(in s)}
        && \multicolumn{1}{c}{(in s)} && \multicolumn{1}{c}{(in GB)}
        \\
    \midrule
        LDSpider (T8) &&
            0.00 &&      67 && 
            -- && -- && -- && -- \\
        LDSpider (T16) &&
            0.00 &&     73 && 
            -- && -- && -- && -- \\
        LDSpider (T32) &&
            0.00 &&     74 && 
             -- && -- && -- && -- \\
        LDSpider (T1,FS) &&
            0.31 &&  1\,798 && 
            1.00 &&  2\,031 &&    320.0 &&  1.2 \\ 
        LDSpider (T8,FS) &&
            0.30 &&  1\,792 && 
            1.00 &&  2\,295 &&    365.9 &&  2.8 \\ 
        LDSpider (T16,FS) &&
            0.31 &&  1\,858 && 
            1.00 &&  1\,945 &&    345.4 &&  1.6 \\ 
        LDSpider (T32,FS) &&
            0.31 &&  1\,847 && 
            1.00 &&  2\,635 &&    588.7 &&  2.6 \\ 
        LDSpider (T32,FS,LBS) &&
            0.03 &&      66 && 
            0.54 &&     765 &&    182.1 &&  7.5 \\ 
        \midrule
        Squirrel (W1) &&
            0.98 &&  6\,663 && 
            1.00 && 11\,821 &&    991.3 &&  3.9 \\ 
        Squirrel (W3) &&
            0.98 &&  2\,686 && 
            1.00 &&  4\,100 &&    681.4 &&  8.6 \\ 
        Squirrel (W9) &&
            0.98 &&  1\,412 && 
            1.00 &&  1\,591 &&    464.8 && 18.1 \\ 
        Squirrel (W18) &&
            0.97 &&  1\,551 && 
            1.00 &&  1\,091 &&    279.8 && 22.1 \\ 
    \bottomrule
    \end{tabular}
\end{table*}

\subsection{Efficiency evaluation}

The second experiment focuses on the efficiency of the crawler implementations. For this purpose, a synthetic Web comprising 200 dereferencing nodes is used since they offer to negotiate the RDF serialisation for transferring the data. This ensures that all crawlers can crawl the complete Web. The other parameters remain as before. 
For LDSpider, we use only the FS instances. We expect both crawlers to be able to crawl the complete cloud and that crawler instances with more threads or workers will crawl faster. The results of the experiment are listed in Table~\ref{tab:results}.\footnote{Due to the lack of space we omit the standard deviations of the measures. While repeating the experiments, the measures turned out to be stable with standard deviations of $\sim$2\% for the RAM, $\sim$5\% for runtime and CPU time. The detailed results can be found at \url{https://w3id.org/hobbit/experiments\#1586886425879,1587151926893,1587284972402,1588111671515,1587121394160,1586886364444,1586424067908,1586374166710,1586374133562}.}

\subsection{Robots Exclusion Protocol check}

In the third experiment, we evaluate whether the crawlers follow the rules defined in a server's \texttt{robots.txt} file. To this end, we configure \Orca{} to generate a smaller Web comprising 25 dereferencing nodes. Each of the nodes copies 10\% of its RDF resources and marks the copies disallowed for crawling using the \texttt{disallow} instruction in its \texttt{robots.txt} file. The average number of requested disallowed resources (RDR) is used as metric. Additionally, we define a delay of 10 seconds between two consecutive requests using the \texttt{Crawl-delay} instruction in the same file. The average node degree of the nodes is configured as 5 while the average resource degree is set to 6. We calculate the crawl delay fulfilment (CDF) which we define as the requested delay divided by the average delay measured by the server. Table~\ref{tab:robots-results} shows the results of this experiment.\footnote{The detailed results can be seen at \url{https://w3id.org/hobbit/experiments\#1575626666061,1575592492658,1575592510594}.}

\begin{table}[tb]
    \centering
    \caption{Results for a Data Web with \texttt{robots.txt} files including disallow and crawl-delay rules. CDF = Crawl delay fulfilment; RDR = Requested disallowed resources.}
    \label{tab:robots-results}
    \begin{tabular}{@{}l@{\ \ }c@{\ \ }c@{\ \ }c@{\ \ }c@{\ \ }r@{}}
    \toprule
        \multicolumn{1}{c}{\multirow{2}{*}{\textbf{Crawler}}} & \multicolumn{3}{c}{\textbf{CDF}} & \multicolumn{1}{c}{\multirow{2}{*}{\textbf{RDR}}} & \multicolumn{1}{c}{\textbf{Runtime}} \\
    \cmidrule{2-4}
         & Min & Max & Avg & & \multicolumn{1}{c}{(in s)} \\
    \midrule
        LDSpider (T32,FS,BFS) &
            0.052 & 0.122 & 0.089 & 0.0 & 224 \\ 
        LDSpider (T32,FS,LBS) &
            0.002 & 0.007 & 0.004 & 0.0 & 43 \\ 
        Squirrel (W18) &
            0.697 & 0.704 & 0.699 & 0.0 & 2384 \\ 
    \bottomrule
    \end{tabular}
\end{table}

\section{Discussion}
\label{sec:discussion}

The experiment results give several insights. As expected, none of the instances of LDSpider were able to crawl the complete synthetic Linked Data Web during the first experiment. Apart from the expected reasons previously mentioned (i.e., the missing support for SPARQL, CKAN nodes and compressed dump files), we encountered two additional issues. First, as mentioned in Section~\ref{sec:benchmarkedCrawlers}, the SPARQL client of LDSpider did 
not store all the crawled triples in the provided triple store. This leads to the different recall values of the LDSpider instances with and without the "FS" extension. Second, although we tried several content handler modules and configurations, LDSpider did not crawl dump files provided as N-Triples. In comparison, the Squirrel instances crawl the complete cloud, except for some triples of RDFa and CKAN nodes.

The second experiment reveals that overall, LDSpider is more resource-efficient than Squirrel. In nearly all cases, LDSpider crawls the Web faster and uses less resources than the Squirrel instances. Only with 9 or more workers Squirrel is able to crawl faster. For the size of the graph, the number of threads used by LDSpider do not seem to play a major role when employing the BFS strategy. It could be assumed that the synthetic Web, with 200 nodes, provides only rare situations in which several nodes are crawled by LDSpider in parallel. However, this assumption can be refuted since Squirrel achieves lower runtimes. Therefore, the load-balancing strategy of Squirrel seems to allow faster crawling of the Web than the BFS of LDSpider. However, the LDSpider (T32,FS,LBS) instance implementing a similar load-balancing strategy aborts the crawling process very early in all three experiments. Therefore, a clearer comparison of both strategies is not possible.

The third experiment shows that both crawlers follow the Robots Exclusion Protocol as both did not request disallowed resources. However, Squirrel seems to insert delays between it's requests---although it reaches on average only 69.9\% of the delay the server asked for---
while LDSpider does not seem take the \texttt{Crawl-delay} instruction into account.

\section{Conclusion}
\label{sec:conclusion}
In this paper, we present \Orca{}---the first extensible FAIR benchmark for Data Web crawlers, which measures the efficiency and effectiveness of crawlers in a comparable and repeatable way. Using \Orca{}, we compared two Data Web crawlers in a repeatable setup. We showed that \Orca{} revealed strengths and limitations of both crawlers. Additionally, we showed that \Orca{} can be used to evaluate the politeness of a crawler, i.e., whether it abides by the Robots Exclusion Protocol. Our approach will be extended in various ways in future work. 
First, we will include HTML pages with Microdata, Microformat or JSON-LD into the benchmark. 
A similar extension will be the addition of further compression algorithms to the dump nodes (e.g., \texttt{tar}), as well as the HDT serialization~\cite{hdt}. 
The generation step will be further improved by adding literals and blank nodes to the generated RDF KGs and altering the dataset sizes. A simulation of network errors will round up the next version of the benchmark.

\section*{Acknowledgements}
This work has been supported by the German Federal Ministry of Transport and Digital Infrastructure (BMVI) within the projects OPAL (19F2028A) and LIMBO (19F2029I).

\bibliographystyle{splncs03}
\bibliography{references}

\end{document}